\documentstyle[11pt,aasms4]{article}

\newcommand{\etal}{~et~al.}

\newcommand{\wisk}[1]{\ifmmode{#1}\else{$#1$}\fi}

\begin{document}

\vspace{-2.0truecm}
\begin{flushright}
KSUPT-97/3, KUNS-1459 \hspace{0.5truecm} August 1997
\end{flushright}
\vspace{-0.5truecm}

\title{Using White Dish CMB Anisotropy Data to Probe Open
  and Flat-$\Lambda$ CDM Cosmogonies}

\author{
  Bharat~Ratra\altaffilmark{1},
  Ken~Ganga\altaffilmark{2},  
  Naoshi~Sugiyama\altaffilmark{3},
  G.~S.~Tucker\altaffilmark{4},
  G.~S.~Griffin\altaffilmark{5},
  H.~T.~Nguy\^en\altaffilmark{6},
  and  
  J.~B.~Peterson\altaffilmark{5}
  }

\altaffiltext{1}{Department of Physics, Kansas State University,
                 Manhattan, KS 66506.}
\altaffiltext{2}{IPAC, MS 100--22, California Institute of Technology, 
                 Pasadena, CA 91125.}    
\altaffiltext{3}{Department of Physics, Kyoto University, 
                 Kitashirakawa-Oiwakecho, Sakyo-ku, Kyoto 606.}
\altaffiltext{4}{Department of Physics, Brown University, Box 1843,
                 Providence, RI 02912.} 
\altaffiltext{5}{Physics Department, Carnegie Mellon University, 5000
                 Forbes Avenue, Pittsburgh, PA 15213.}
\altaffiltext{6}{Jet Propulsion Laboratory, California Institute of
                 Technology, Pasadena, CA 91109.}

\begin{abstract}
  We use data from the White Dish experiment to set limits on cosmic
  microwave background radiation anisotropies in open and 
  spatially-flat-$\Lambda$ cold dark matter cosmogonies.
  We account for the White Dish calibration uncertainty, and marginalize
  over the offset and gradient removed from the data. Our
  2-$\sigma$ upper limits are larger than those derived previously. These
  upper limits are consistent with those derived from the $COBE$-DMR 
  data for all models tested.
\end{abstract}

\keywords{cosmic microwave background---cosmology: observations---large-scale
  structure of the universe}

\section{Introduction}

Ganga et al. (1997a, hereafter GRGS) developed techniques to account for 
uncertainties, such as those in the beamwidth and calibration, in likelihood
analyses of cosmic microwave background (CMB) anisotropy data. GRGS and Ganga
et al. (1997b,1998) used these techniques, in conjunction with theoretical CMB 
anisotropy spectra, in analyses of the UCSB South Pole 1994 (Gundersen et al. 
1995), the SuZIE (Church et al. 1997), and the MAX 4+5 (Tanaka et al. 1996; 
Lim et al. 1996, and references therein) CMB anisotropy data sets. Bond \&\ 
Jaffe (1997) have also analyzed the UCSB South Pole 1994 data and the 
Saskatoon (Netterfield et al. 1997) data.

In this paper we present a similar analysis of the Tucker et al. (1993, 
hereafter T93) White Dish CMB anisotropy data collected at the South Pole. The 
White Dish detector and telescope are described in Tucker et al. (1994).
Data were taken in a frequency band centered at 90 GHz. The FWHM of the beam,
assumed to be gaussian, is $12^\prime$. Five interlocked circles on the sky,
centered at constant elevation and declination, were observed. The circle
centers
are separated by $15^\prime$ on the sky, and each circle intersects at least
one neighbour at two points. The circles, of diameter $28^\prime$ on the 
sky, were sampled at 128 equally spaced points. Griffin et al. (1997) 
describe the full White Dish data set.

T93 analyze a small subset of the White Dish data in two different ways,
which they refer to as Method I and Method II. They consider only the set 
of points, at two 
different elevations, defined by where the interlocking circles intersect.
Method I uses two sets of two-beam temperature differences (one set at each 
elevation). Method II uses a single set of ``quadrupole" temperature 
differences, obtained by appropriately combining the corresponding two-beam 
differences at each elevation. Further details are given in T93. 

Neither method results in a 2-$\sigma$ detection of CMB anisotropy (T93).
Since Method II provides a less restrictive upper limit than does Method I, we
do not record Method II results here. Method I and Method II use essentially
the same CMB anisotropy observations; they thus can not be combined to 
provide a tighter upper limit. T93 remove an offset and linear gradient 
from each Method I scan, prior to binning in right ascension.

T93 and Tucker et al. (1994) describe how White Dish was calibrated. The 
absolute calibration uncertainty is 30\% (1-$\sigma$).

In $\S$2 we briefly summarize the computational techniques used in our 
analysis. See GRGS for further details. Results are presented and discussed in 
$\S$3, and conclusions are given in $\S$4.

\section{Summary of Computation}

Figure 1 shows the reduced White Dish Method I data. 

For two White Dish circles centered at azimuth angles $\phi_i$ and $\phi_j$, 
the Method I window function is 
\begin{eqnarray}
 W_{lij} 
     &=& e^{-\sigma_{\rm G}{}^2 (l + 0.5)^2} \times \nonumber \\
     &\bigg[& 2 P_l\left({\rm cos}\left\{\sqrt{\Delta\theta^2 + (\phi_i - 
        \phi_j)^2}\right\}\right) \nonumber \\
     & & - P_l\left({\rm cos}\left\{\sqrt{\Delta\theta^2 + (\phi_i - 
        \phi_j - \Delta\phi)^2}\right\}\right)
        - P_l\left({\rm cos}\left\{\sqrt{\Delta\theta^2 + (\phi_i - \phi_j
        + \Delta\phi)^2}\right\}\right) \bigg]. 
\end{eqnarray}
Here the gaussian beamwidth $\sigma_{\rm G} = 12^\prime /\sqrt{8 {\rm ln} 2}$, 
$P_l$ is a Legendre polynomial of order $l$ (the multipole), the throw 
$\Delta\phi = 15^\prime$, and the separation in elevation 
$\Delta\theta$ is zero for two points at the same elevation and is 
$23.6^\prime$  for two points at 
different elevations.

We do not record results from the Method II analysis here. However, since the 
Method II window function has not been given in the published literature, we 
note that it is 
\begin{eqnarray}
  W_{lij} 
      &=& e^{-\sigma_{\rm G}{}^2 (l + 0.5)^2} \times \nonumber \\
      &\bigg[& 2 P_l\left({\rm cos}\left\{\phi_i - \phi_j \right\}\right) 
        \nonumber \\
      & & - 0.5 P_l \left({\rm cos}\left\{\phi_i - \phi_j - 
        \Delta\phi\right\} \right) 
        - 0.5 P_l \left({\rm cos}\left\{\phi_i - \phi_j + \Delta\phi\right\}
        \right) \nonumber \\
      & & + 0.5 P_l\left({\rm cos}\left\{\sqrt{\Delta\theta^2 + (\phi_i - 
        \phi_j - \Delta\phi)^2}\right\}\right)
        + 0.5 P_l\left({\rm cos}\left\{\sqrt{\Delta\theta^2 + (\phi_i - \phi_j
        + \Delta\phi)^2}\right\}\right) \nonumber \\
      & & - P_l\left({\rm cos}\left\{\sqrt{\Delta\theta^2 + (\phi_i - 
        \phi_j)^2}
        \right\}\right)\bigg] , 
\end{eqnarray}
where the notation is the same as that for equation (1), except $\Delta\theta$
is always 23.6$^\prime$ here. The zero-lag 
parts of the Method I and II window functions are shown in Figure 2 and the
window function parameters are given in Table 1. 

Figure 2 also shows some of the model CMB anisotropy spectra we use.
These spectra are described in Ratra et al. (1997). In addition to the
flat bandpower and fiducial cold dark matter (CDM) model spectra, we consider 
spatially open CDM models as well as spatially flat CDM models with a 
cosmological constant ($\Lambda$). The low-density open and flat-$\Lambda$ 
models are consistent with most current observations. See Bunn \& White (1997),
G\'orski et al. (1998), Gott (1997), Turner (1997), Peacock (1997),
Cole et al. (1997), and Gardner et al. (1997) for discussions.

The computation of the CMB anisotropy spectra is described in Sugiyama (1995).
These computations assume gaussian\footnote
{It has recently been suggested that degree-scale CMB anisotropy
observations might indicate nongaussianity (Gazta\~naga, Fosalba, \& 
Elizalde 1997). In this case, the good fit of gaussian models to the data,
indicated by very low reduced $\chi^2$ values for some models (Ganga, Ratra,
\& Sugiyama 1996, also see Lineweaver \& Barbosa 1997), would be a coincidence.
As would the almost identical conclusions deduced from the UCSB South Pole
1994 and MAX 4+5 data sets (Ganga et al. 1998).}, 
adiabatic primordial energy density power spectra. The open model computations
assume the open-bubble inflation model energy-density spectrum (Ratra \&
Peebles 1994,1995; Bucher, Goldhaber, \& Turok 1995; Yamamoto, Sasaki, \&
Tanaka 1995). The flat-$\Lambda$ model computations assume the scale-invariant
energy-density power spectrum (Harrison 1970; Peebles \& Yu 1970; Zel'dovich
1972).

The CMB anisotropy spectra are parameterized by their quadrupole-moment
amplitude $Q_{\rm rms-PS}$, the clustered-mass density parameter $\Omega_0$,
the baryonic-mass density parameter $\Omega_B$, and the Hubble parameter
$h$ [$= H_0/(100 h\ {\rm km}\ {\rm s}^{-1}\ {\rm Mpc}^{-1})$]. The 
cosmological parameter values tested were chosen on the basis of consistency
with current non-CMB observations (Ratra et al. 1997). Table 2 shows
the parameter values used in our analyses.

Figure 3 shows the moments $(\delta T_{\rm rms}{}^2)_l = T_0{}^2 (2l+1) 
C_l W_l/(4\pi)$ (where $T_0$ is the CMB temperature now, GRGS, eqs. [5] \&
[6]) for the Method I window function and the CMB anisotropy spectra of
Figure 2. These moments show the angular scales where the Method I experiment
is most sensitive, given a CMB anisotropy model (GRGS). Examination
of the White Dish window function by itself (i.e., without reference to a 
CMB anisotropy spectrum) does not give an accurate indication 
of the multipoles to which the experiment is sensitive (Table 1 and Figure 2).

The computation of the likelihood function is described in GRGS. We assume a
uniform prior in the amplitudes of the offset and gradient removed
and marginalize over these amplitudes, i.e., we integrate over all possible 
offset and gradient amplitudes when determining the likelihood function 
(Bond et al. 1991, also see Bunn et al. 1994; GRGS; Church et al. 1997; 
Ganga et al. 1997b). Calibration uncertainty is accounted 
for as described in GRGS. Figure 4 shows the Method I likelihood functions for 
the CMB anisotropy spectra of Figure 2. These likelihood functions account 
for all the above additional uncertainties.

In agreement with the conclusion of T93, there are no 2-$\sigma$ detections
of anisotropy (defined using the prescription given in GRGS). To derive 
$Q_{\rm rms-PS}$ upper limits we assume a uniform prior in $Q_{\rm rms-PS}$ and 
integrate the posterior probability density distribution function starting 
from 0 $\mu$K until 95.5\% of the area is encompassed. This is the 2-$\sigma$ 
highest posterior density (HPD) prescription; see GRGS for further details.
(The corresponding 2-$\sigma$ equal tail limits, e.g., GRGS, are significantly
larger and so not recorded here.) Table 2 gives these $Q_{\rm rms-PS}$
2-$\sigma$ HPD limits, as well as those for bandtemperature $\delta T_l
= \delta T_{\rm rms}/[\sum^\infty_{l=2} [(l+0.5) W_l/\{l(l+1)\}]]^{0.5}$.

\section{Results and Discussion}

From Table 2, the Method I $\delta T_l$ 2-$\sigma$ upper limit for the flat
bandpower angular spectrum is 150 $\mu$K. This accounts for the marginalization
over the amplitudes of the offset and gradient removed, as well as the 
calibration uncertainty. If instead we use a gaussian autocorrelation
function (GACF), with a coherence angle of 0.15$^\circ$, to describe the 
CMB anisotropy, the corresponding bandtemperature $\delta T_l$ limit is 
140 $\mu$K, in good agreement with the flat bandpower angular spectrum 
result. Ignoring the marginalization over
offset and gradient removed, and the calibration uncertainty, the flat 
bandpower $\delta T_l$ 2-$\sigma$ upper limit is 54 $\mu$K. 
Accounting only for the marginalization over offset
and gradient, the flat bandpower $\delta T_l$ 2-$\sigma$ limit rises to
96 $\mu$K. These numerical values should be compared to the T93 bandtemperature
$\delta T_l$ 2-$\sigma$ upper limit of 44 $\mu$K, derived for a GACF with a 
coherence angle of 0.15$^\circ$. The T93 computation 
does not account for calibration uncertainty, nor does it account for the 
marginalization over the amplitudes of the offset and gradient removed 
from the data.

The White Dish Method I limits derived here are larger than that of 
T93. The Method II
upper limits (not recorded here) are larger than those of Method I. This is
in qualitative agreement with the results of T93. For all models tested,
the White Dish $Q_{\rm rms-PS}$ 2-$\sigma$ upper limits derived here are
consistent with the DMR detections (G\'orski et al. 1996,1998; Stompor 1997).

The limits derived depend sensitively on whether offset and gradient removal
are accounted for in the likelihood analysis, and whether calibration 
uncertainty is included. There are situations in which the calibration 
uncertainty need not be included (for example, when considering the ratio of
two different measurements made with the same instrument). For comparison
with other experiments, however, calibration uncertainty must be included.

The variation of the $\delta T_l$ upper limit from model to model gives an 
indication of the accuracy of the flat bandpower approximation. From Table 
2 we see that there is a difference of $\sim 20\%$ between the two extreme
cases. This is comparable to the $\sim 25\%$ difference for SuZIE (Ganga
et al. 1997b), and larger than the $\sim 10\%$ variation for UCSB South Pole
1994 and MAX 4+5 (GRGS; Ganga et al. 1998).

\section{Conclusion}

In our likelihood analyses of the White Dish Method I CMB anisotropy
data we have marginalized over the amplitudes of the offset and gradient
removed from the data, and have explicitly accounted for calibration 
uncertainty. There are no 2-$\sigma$ detections of anisotropy for the models 
tested, in agreement with the conclusion of T93. As a consequence of the 
additional effects accounted for here, the limits we have derived are 
less restrictive than those derived by T93. These limits are 
consistent with the $COBE$-DMR anisotropy amplitudes for the models tested.
Hence, contrary to earlier assertions (e.g., Ostriker \& Steinhardt 1995; 
Ratra et al. 1997), the T93 White Dish data subset does not seriously 
constrain cosmological parameter values for reasonable DMR-normalized models.
%

\bigskip

We acknowledge helpful discussions with J. Gundersen, L. Page, and G. Rocha. 
BR acknowledges support from NSF grant EPS-9550487 with matching support from 
the state of Kansas and from a K$^*$STAR First award. This work was partially 
carried out at the Infrared Processing and Analysis Center and the Jet 
Propulsion Laboratory of the California Institute of Technology, under a 
contract with the National Aeronautics and Space Administration.

\clearpage

\begin{table}
\begin{center}
\caption{Numerical Values for the Zero-Lag Window Function 
Parameters\tablenotemark{a}}
\vspace{0.3truecm}
\tablenotetext{{\rm a}}{The value of $l$ where $W_l$ is
largest, $l_{\rm m}$, the two values of $l$ where $W_{l_{e^{-0.5}}} =
e^{-0.5} W_{l_{\rm m}}$, $l_{e^{-0.5}}$, the effective multipole,
$l_{\rm e} = I(lW_l)/I(W_l)$, and 
$I(W_l) = \sum^\infty_{l=2}(l+0.5)W_l/\{l(l+1)\}$.}
\begin{tabular}{lccccc}
\tableline\tableline
  & $l_{e^{-0.5}}$ & $l_{\rm e}$ & $l_{\rm m}$
  & $l_{e^{-0.5}}$ & $\sqrt{I(W_l)}$  \\
\tableline
Method I  & 297 & 477 & 539 & 825 & 1.18 \\
Method II & 415 & 579 & 615 & 833 & 0.725 \\
\tableline
\end{tabular}
\end{center}
\end{table}

\begin{table}
\renewcommand{\arraystretch}{0.9}
\begin{center}
\caption{Upper Limits\tablenotemark{a} $\ $on $Q_{\rm rms-PS}$ and 
$\delta T_l$ (in $\mu$K) from the Method I Analysis}
\vspace{0.3truecm}
\tablenotetext{{\rm a}}{2-$\sigma$ (95.5\%) HPD limits.}
\begin{tabular}{lccc}
\tableline\tableline
$\#$ & $(\Omega_0 , h , \Omega_B h^2)$ & $Q_{\rm rms-PS}$ & $\delta T_l$ \\
\tableline
         O1 &(0.1, 0.75, 0.0125)& 72 & 150 \\
         O2 &(0.2, 0.65, 0.0175)& 70 & 150 \\
         O3 &(0.2, 0.70, 0.0125)& 75 & 150 \\
         O4 &(0.2, 0.75, 0.0075)& 81 & 150 \\
         O5 &(0.3, 0.60, 0.0175)& 66 & 160 \\
         O6 &(0.3, 0.65, 0.0125)& 72 & 160 \\
         O7 &(0.3, 0.70, 0.0075)& 77 & 150 \\
         O8 &(0.4, 0.60, 0.0175)& 63 & 160 \\
         O9 &(0.4, 0.65, 0.0125)& 67 & 160 \\
        O10 &(0.4, 0.70, 0.0075)& 71 & 160 \\
        O11 &(0.5, 0.55, 0.0175)& 54 & 170 \\
        O12 &(0.5, 0.60, 0.0125)& 57 & 170 \\
        O13 &(0.5, 0.65, 0.0075)& 61 & 160 \\
        O14 &(1.0, 0.50, 0.0125)& 59 & 170 \\
 $\Lambda$1 &(0.1, 0.90, 0.0125)& 71 & 170 \\
 $\Lambda$2 &(0.2, 0.80, 0.0075)& 68 & 170 \\
 $\Lambda$3 &(0.2, 0.75, 0.0125)& 64 & 170 \\
 $\Lambda$4 &(0.2, 0.70, 0.0175)& 61 & 170 \\
 $\Lambda$5 &(0.3, 0.70, 0.0075)& 63 & 170 \\
 $\Lambda$6 &(0.3, 0.65, 0.0125)& 59 & 170 \\
 $\Lambda$7 &(0.3, 0.60, 0.0175)& 56 & 180 \\
 $\Lambda$8 &(0.4, 0.65, 0.0075)& 61 & 170 \\
 $\Lambda$9 &(0.4, 0.60, 0.0125)& 57 & 170 \\
$\Lambda$10 &(0.4, 0.55, 0.0175)& 54 & 180 \\
$\Lambda$11 &(0.5, 0.60, 0.0125)& 57 & 170 \\
  Flat      &         ...       & 97 & 150 \\
\tableline
\end{tabular}
\end{center}
\end{table}


\clearpage
\centerline{\bf Figure Captions}

\begin{figure}[htbp]
  \caption{Measured thermodynamic temperature differences (with $\pm 1$-$\sigma$
    error bars) on the sky as a 
    function of scan position. The triangles correspond to the $+$ elevation
    scans of T93 and the circles correspond to the $-$ elevation scans.}
  \label{fig:dataplot}
\end{figure}
\begin{figure}[htbp]
  \caption{CMB anisotropy multipole moments $l(l+1)C_l/(2\pi )\times
    10^{10}$ (broken lines, scale on left axis) as a function of
    multipole $l$, for selected open models O2, O12, and O14 (fiducial CDM),
    flat-$\Lambda$ models $\Lambda$4, $\Lambda$11, and Flat (bandpower), 
    normalized to the DMR maps
    (G\'orski et al. 1996,1998; Stompor 1997). See Table 2 for model-parameter
    values. Also shown are the White Dish Method I and II zero-lag
    window functions $W_l$ (solid lines, scale on right axis). See Table 1
    for $W_l$-parameter values.}
  \label{fig:win_mod}
\end{figure}
\begin{figure}[htbp]
  \caption{$(\delta T_{\rm rms}{}^2)_l$ as a function of $l$, for the 
    Method I window function, and for the selected model spectra shown in 
    Figure 2. These curves should be compared to the Method I window function 
    shown in Figure 2. Note the multiple ``sensitivity" peaks for some of the 
    spectra. Note also that these peaks correspond to a different angular 
    scale in each of the models.}
  \label{fig:trms}
\end{figure}
\begin{figure}[htbp]
  \caption{Method I likelihood functions for the six CMB anisotropy spectra
    of Figure 2.} 
  \label{fig:likeplot}
\end{figure}

\clearpage
\pagestyle{empty}

\begin{center}
  \leavevmode
  \epsfxsize=5.5truein
  \epsfbox{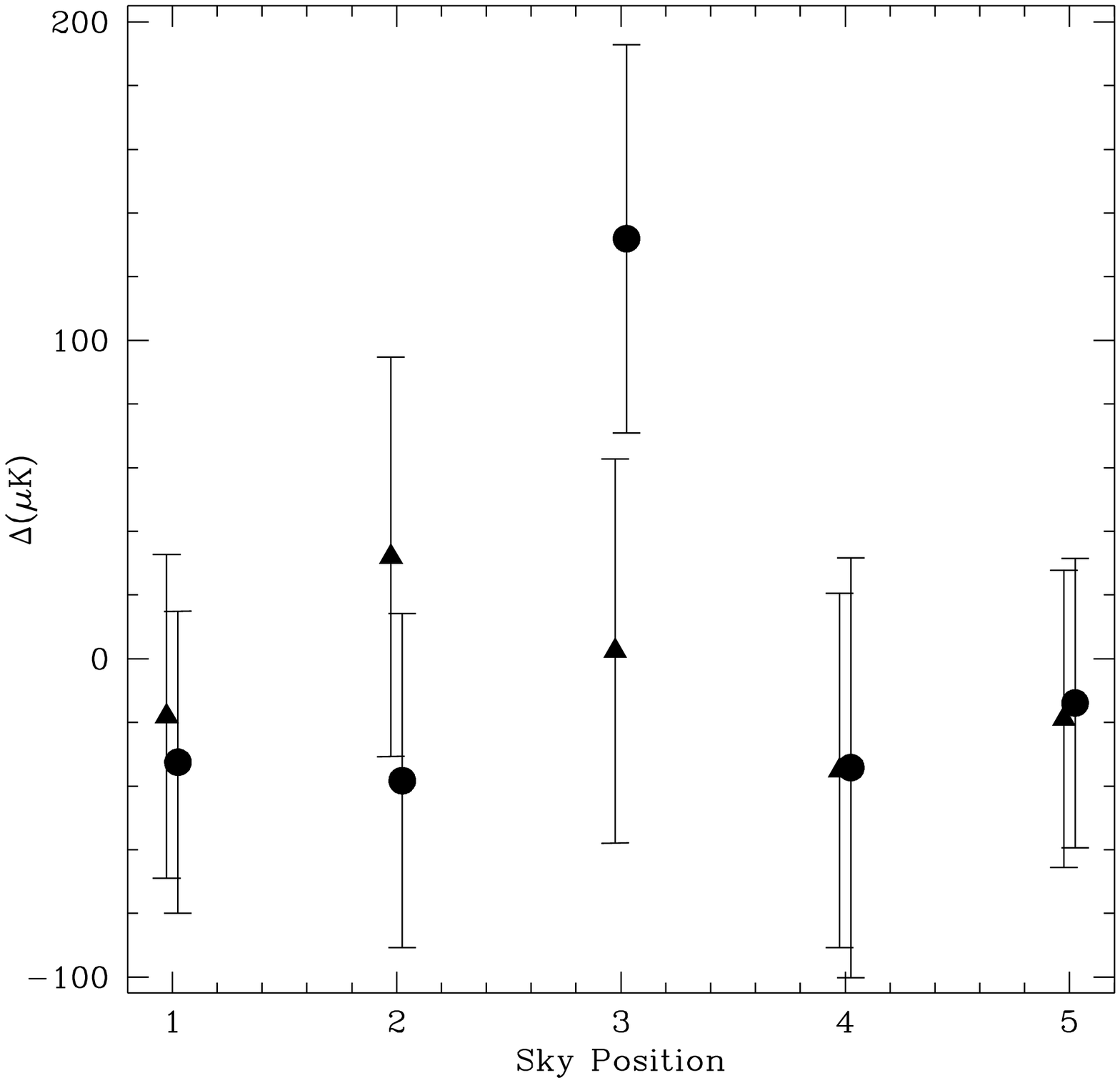}
\end{center}  
\vfill
Figure~\ref{fig:dataplot}

\clearpage
\begin{center}
  \leavevmode
  \epsfxsize=5.5truein
  \epsfbox{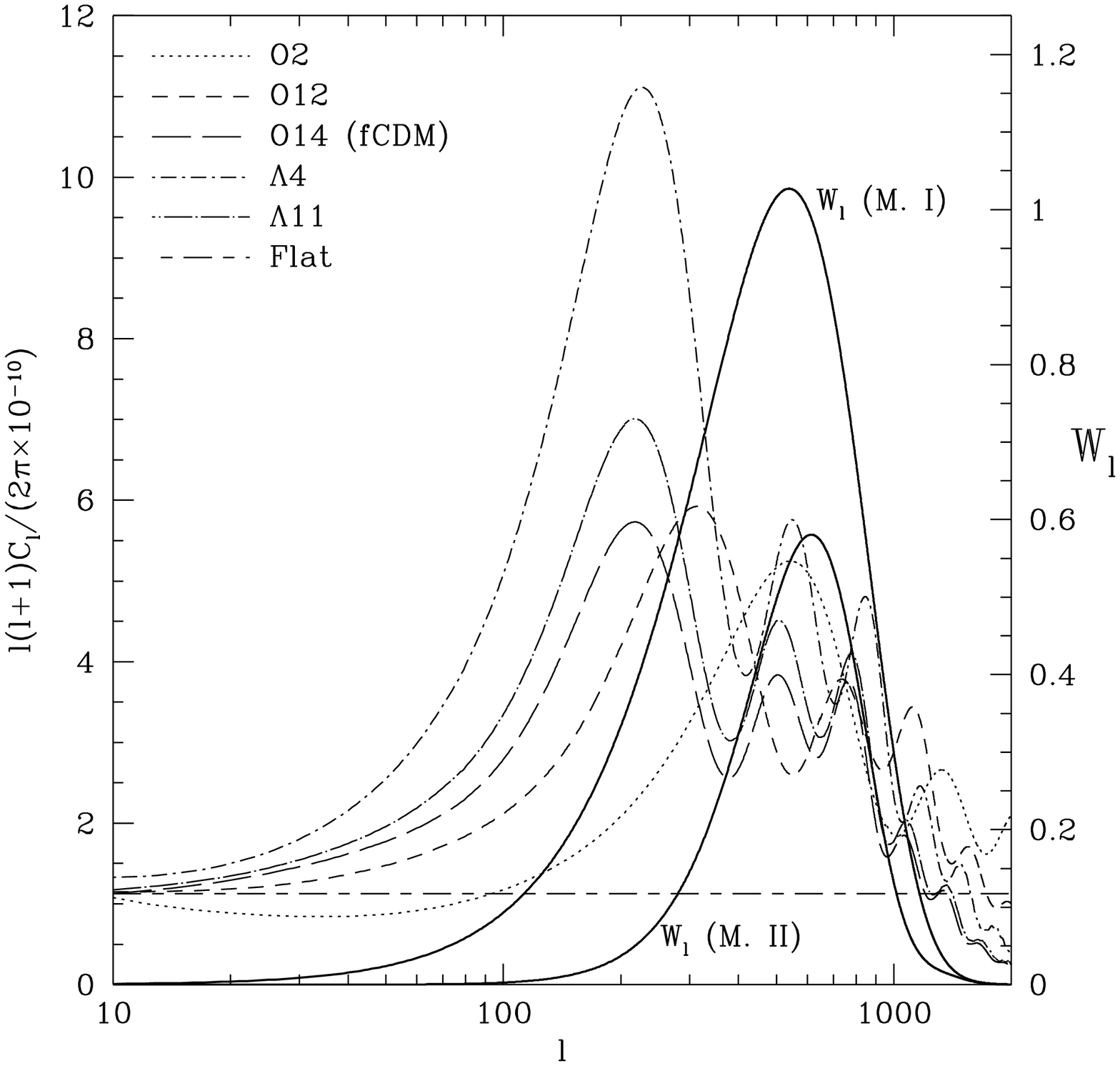}
\end{center}  
\vfill
Figure~\ref{fig:win_mod}

\clearpage
\begin{center}
  \leavevmode
  \epsfxsize=5.5truein
  \epsfbox{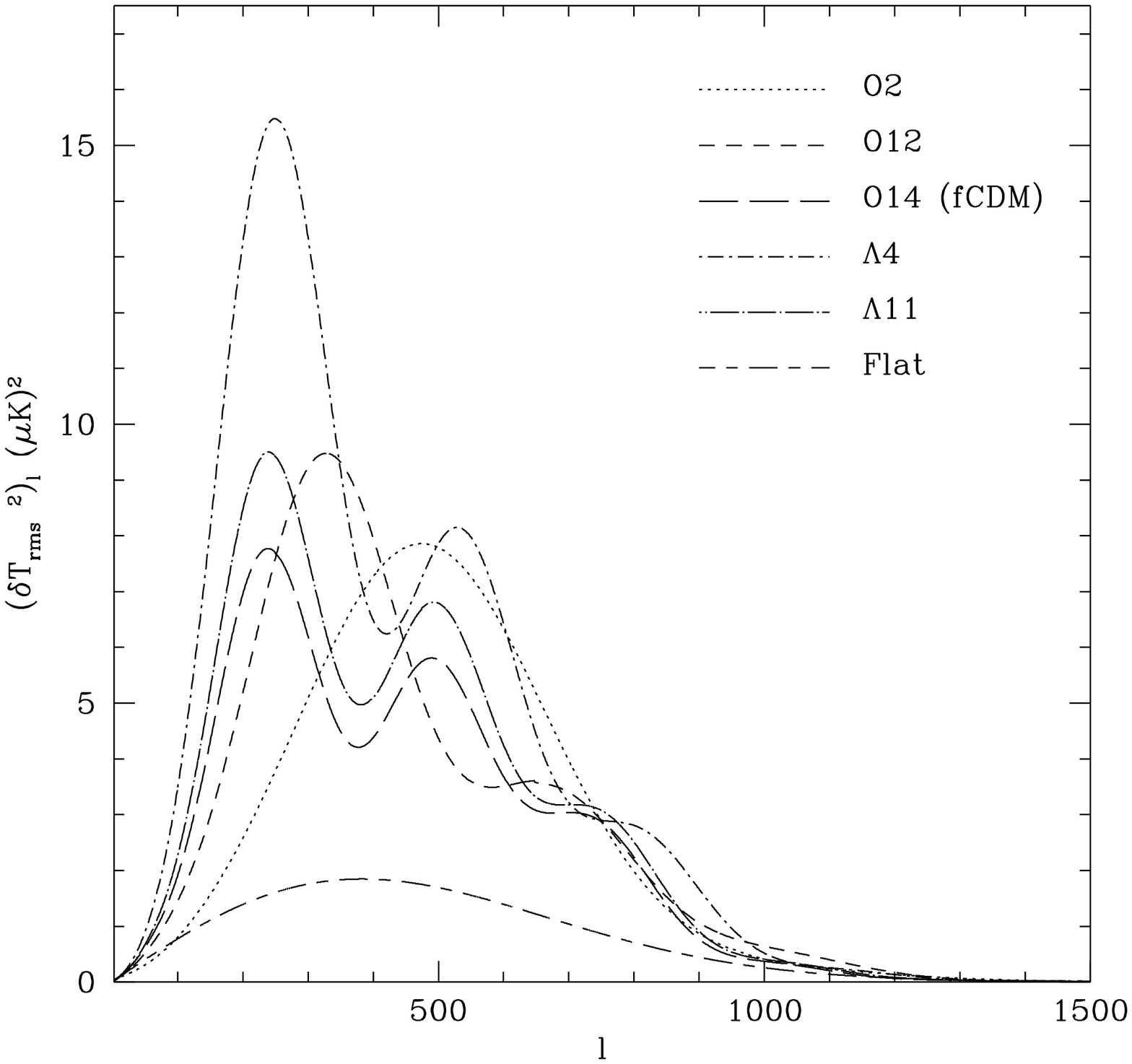}
\end{center}  
\vfill
Figure~\ref{fig:trms}

\clearpage
\begin{center}
  \leavevmode
  \epsfxsize=5.5truein
  \epsfbox{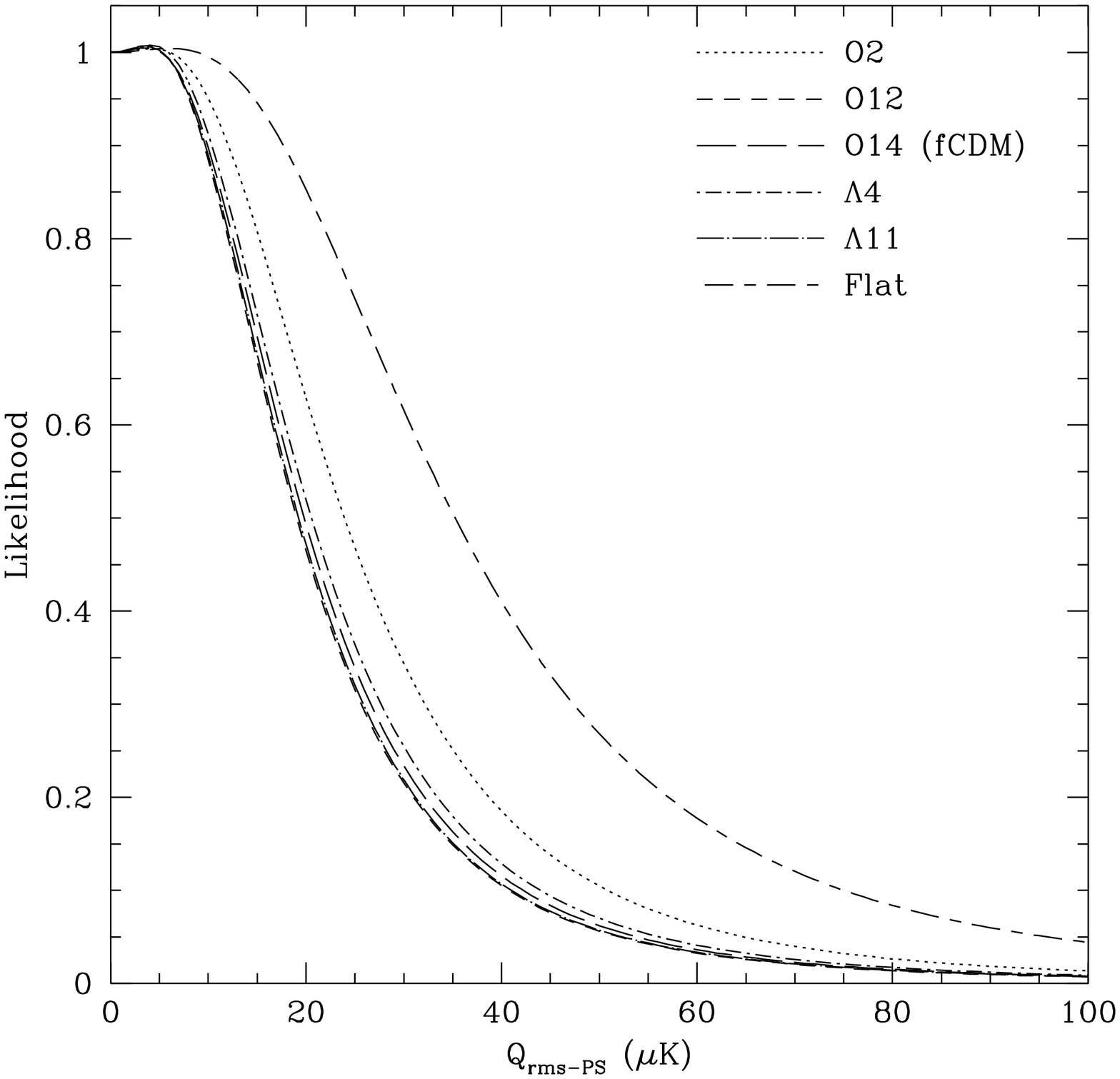}
\end{center}  
\vfill
Figure~\ref{fig:likeplot}

\end{document}